\documentclass[aps,prb]{revtex4}
\usepackage{amsmath,amsfonts,graphicx}

\newcommand{\xx}{\mathbf{r}}

\begin{document}

\title{Non-perturbative phenomena in semiconductor four-wave mixing spectra}

\author{Mikhail Erementchouk}
\author{Michael N. Leuenberger}\email{mleuenbe@mail.ucf.edu}
\affiliation{NanoScience Technology Center and Department of Physics, University of Central
Florida, Orlando, FL 32826}

\author{L. J. Sham}
\affiliation{Department of Physics, University of California San Diego, La
Jolla, CA 92093-0319}

\begin{abstract}
Non-perturbative phenomena in four-wave mixing spectra of semiconductors
are studied using the exact solution of a widely used phenomenological
non-linear equation of motion of the exciton polarization. It is shown
that Coulomb interaction, included in the nonlinearity, leads to two
characteristic effects, which are essentially of dynamical origin,
--- a split of the exciton peak and a non-monotonous dependence of
the response at the exciton frequency on the magnitude of the external
field. Relations between the spectral features and the parameters of the
system is obtained. It is found that the transition from perturbative to
non-perturbative regimes is controlled by parameters inversely
proportional to the decay rate. It implies that the condition of low
excitation density does not necessarily warrant applicability of the
perturbational approach.
\end{abstract}


\maketitle

\section{Introduction}

One of the clearest manifestations of the many-body effects in
semiconductors is the the phenomenon of the non-linear optical response. The
Pauli blocking and the Coulomb interaction between the quasiparticles lead
to the dependence of the polarization dynamics on its spatial
distribution.\cite{BINDER:1995,OSTREICH:1994,OSTREICH:1998,CHERNYAK:1998,CIUTI:2000,SAVASTA:2001,TAKAYAMA:2002}
When a semiconductor quantum well is excited by two successive pulses, they
produce the signal in directions which are prohibited in the linear regime
while still conserve momentum in four-wave and multi-wave mixing.
The dynamical
origin\cite{WEGENER:1990} of the formation of the four-wave mixing signal
is naturally incorporated into
the description in terms of the exciton 
modes characterized by the frequency $\omega_0$ and the in-plane wave
vector $\mathbf{k}$. From this perspective the effects of the Pauli
blocking and of the Coulomb interaction are clearly different. The Pauli
blocking reduces \textit{locally} the intensity of the field-matter
interaction according to the magnitude of the local polarization. As a
result, the excitation field with particular value of the wave-vector
$\mathbf{q}$ becomes
coupled to the 
exciton modes with different $\mathbf{k}\ne \mathbf{q}$. The Coulomb
interaction between the excitons leads to the direct coupling between
the 
exciton modes. In particular, when two modes with $\mathbf{k}_1$ and
$\mathbf{k}_2$ are excited, the relevant coupled modes are characterized
by $\mathbf{k}^{(n)} = \mathbf{k}_2 + n\Delta\mathbf{k}$, where $-\infty <
n < \infty$ is an integer number and $\Delta\mathbf{k} = \mathbf{k}_1 -
\mathbf{k}_2$. Although initially all energy is concentrated in the modes
$\mathbf{k}^{(0)}$ and $\mathbf{k}^{(1)}$ during the evolution the energy
is redistributed among the coupled modes. In particular it leads to the
formation of the four-wave mixing (FWM) signals, which correspond to $n=2$
and $n = -1$.

It follows from this picture that the redistribution of the energy between
the modes characterized by different $\mathbf{k}$ should not be the only
manifestation of the mode coupling. This coupling should also lead to
the modification of the frequencies of the 
exciton modes. Indeed, without the interaction and neglecting the
dispersion of the exciton modes one has a many-fold degeneracy at the
exciton resonance frequency $\omega_0$. The interaction between the modes
should lift the degeneracy and, if the excitation is sufficiently strong,
make the split of the exciton peak visible in the FWM spectrum.

\section{The nonlinear dynamics of the exciton polarization}

The effect of the modification of the frequencies of the 
exciton modes caused by the coupling is a nonperturbative
effect.\cite{SHACKLETTE:2002,WUHR:2004} In order to study the problem we
present the dynamics of the exciton polarization as the exact solution to
the phenomenological nonlinear equation\cite{SCHAFER:1996}
\begin{equation}\label{eq:eq_of_motion}
\begin{split}
  \dot{P}(t; \xx) = -(i\omega_0 + \gamma)& P(t; \xx) - i \beta |P(t; \xx)|^2P(t; \xx) \\
  -i & E(t; \xx)\left[1 - |P(t; \xx)|^2/P_{sat}^2\right],
\end{split}
\end{equation}
where $\omega_0$ is the detuning, i.e. the difference between the frequency
of the external field and the exciton frequency in the stationary frame,
$\gamma$ is the phenomenological decay rate, $E(t; \xx)$ is the envelope
magnitude of the external field, $P_{sat}^2$ is the exciton saturation
density and, finally, $\beta = \beta' - i \beta''$, with $\beta', \beta''
\geq 0$, is a phenomenological parameter quantifying the interaction between
the excitons. The positive real and negative imaginary parts of this
parameter constitute the excitation induced shift (EIS) and the excitation
induced decay (EID), respectively. This phenomenological equation has
provided valuable interpretations for nonlinear
measurements.\cite{SCHAFER:1996,KNER:1997,FU:1997}. The terms may be viewed
as the short-time limit of the memory function,\cite{OSTREICH:1998} which is
exact to the third order in the exciting electric field.\cite{AXT:1994} The
nonpertubative solutions have been considered by
Refs.~\onlinecite{KWONG:2001,KNER:1999} and many papers referenced in
Ref.~\onlinecite{WUHR:2004}.

\begin{figure}
  \includegraphics[width=3 in]{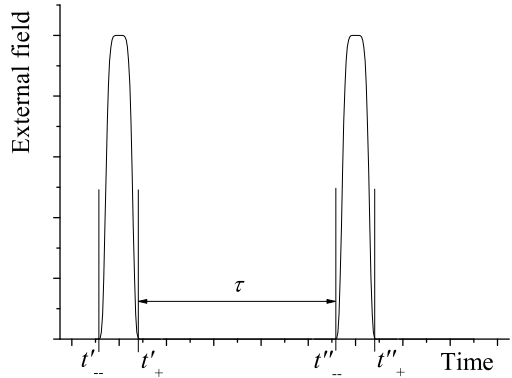}\\
  \caption{The semiconductor is excited by two short pulses separated by the delay time
  $\tau = t_-''-t_+'$.}\label{fig:excitation_setup}
\end{figure}

We consider the excitation of the semiconductor by two short pulses (see
Fig.~\ref{fig:excitation_setup}) acting on the system during the intervals
$t_-' < t < t_+'$ and $t_-'' < t < t_+''$, respectively. Between the
pulses, $t_+' < t < t_-''$ and after the second pulse $t > t_+''$ the
dynamics of the exciton polarization is free and satisfies
Eq.~(\ref{eq:eq_of_motion}) with $E=0$. We study the free dynamics
polarization $P_0(t)$ before calculating the polarization $P_{1,2}(t)$ of
the driven dynamics. Then we are going to solve for the polarization in
the following order: $P_1(t)$ between $t_-' < t < t_+'$, $P_{10}(t)$
between $t_+' < t < t_-''$, $P_2(t)$ between $t_-'' < t < t_+''$, and at
the end the final solution $P_{20}(t)$ for $t
> t_+''$.

The free dynamics [$E(t)=0$] can be solved exactly by noting that
Eq.~\eqref{eq:eq_of_motion} in this regime is reduced to the simple form
\begin{equation}\label{eq:effective_linear}
  \dot{P_0}(t;\xx) = -\left[i\omega_0 + \gamma + i\Omega(t;\xx)\right]P_0(t;\xx),
\end{equation}
where $\Omega(t;\xx) = \beta|P_0(t; \xx)|^2$, and by observing that
Eq.~\eqref{eq:eq_of_motion} yields an equation for the magnitude of the
polarization
\begin{equation}\label{eq:p_amp_equation}
  \frac{d}{dt}|P_0(t; \xx)|^2 = -2[\gamma + \beta'' |P_0(t; \xx)|^2]|P_0(t;
  \xx)|^2
\end{equation}
with the solution
\begin{equation}\label{eq:p_amp_solution}
 |P_0(t; \xx)| = |P_0(0; \xx)| e^{-\gamma t} A(t; |P_0(0,\mathbf{r})|),
\end{equation}
where
\begin{equation}\label{eq:amplitude_factor}
  A^2(t; |P_0(0,\mathbf{r})|) = \left[1+ \frac{\beta''|P_0(0; \xx)|^{2}}{\gamma}
  \left(1 - e^{-2\gamma t}\right)\right]^{-1}.
\end{equation}
The free propagator may be expressed in terms of the amplitude modulation
$A(t;\mathbf{r})$ and the phase modulation $\varphi(t;\mathbf{r})$,
\begin{equation}
   P_0(t;\mathbf{r})/ P_0(0;\mathbf{r})  =: \theta\left(t; |P_0(0,\mathbf{r})|\right)
   = A(t;|P_0(0,\mathbf{r})|)  e^{-i\omega_0 t -\gamma t + i \varphi(t;\mathbf{r})},
\end{equation}
where
\begin{equation}\label{eq:phase_shift}
  \varphi(t;\mathbf{r}) = \frac{\beta'}{\beta''} \ln A\left(t;|P_0(0,\mathbf{r})|\right).
\end{equation}
The amplitude term shows the nonpertubative effect of EID. It is interesting
to note that EID does not lead to a mere modification of the decay rate,
$\gamma$. Instead the amplitude modulation $A(t;\mathbf{r})$ decreases to  a
fraction of the initial amplitude at twice the linear rate, i.e., $2\gamma$.
The phase modulation shows an oscillation dependent on the nonlinear quality
factor $\beta'/\beta''$. This is related to the Goldstone mode in the
excitons studied in Ref.~\onlinecite{OSTREICH:1999}.


The initial conditions for Eq.~(\ref{eq:effective_linear}) are determined
by the polarization distribution right after the external field is
switched off. 
We find the immediate response of the system assuming that the duration of
the excitation pulses is much shorter than the typical dynamical time
scales determined by detuning and the decay rate.

We turn now to the driven ($E(t)\ne 0$) time evolution of the polarization,
i.e. we consider the time interval $t_- < t < t_+$ where the particular
excitation pulse does not vanish and factor out the term
$\exp(i\mathbf{k}\cdot\xx)$ so that we can consider the excitation pulse to
be spatially homogeneous. Neglecting the contribution to the phase $\sim
\int_{t_-}^{t_+}\left[i\omega_0 + \gamma +
i\Omega(t;\xx)\right]P_{1,2}(t;\xx) dt$ we solve the dynamical equation and
find the relation between the polarization at the instances $t_-$ and $t_+$
\begin{widetext}
\begin{equation}\label{eq:immediate response}
  P_{1,2}(t_+) = {P'}_{1,2}(t_-) - i
  \sqrt{P_{sat}^2-{P'}_{1,2}^2(t_-)}\tanh\left\{\sqrt{P_{sat}^2-{P'}_{1,2}^2(t_-)}\frac{{\epsilon}}{P_{sat}^2}
  -\mathrm{atanh}\left[\frac{{P''}_{1,2}(t_-)}{\sqrt{P_{sat}^2-{P'}_{1,2}^2(t_-)}}\right]\right\},
\end{equation}
where ${P'}_{1,2}(t_-) = \mathrm{Re}[P_{1,2}(t_-)]$ and ${P''}_{1,2}(t_-) =
\mathrm{Im}[P_{1,2}(t_-)]$, and we have introduced the area of the exciting
pulse $\epsilon = \int dt E(t)$.
\end{widetext}

\begin{figure*}
  \includegraphics[width=6in]{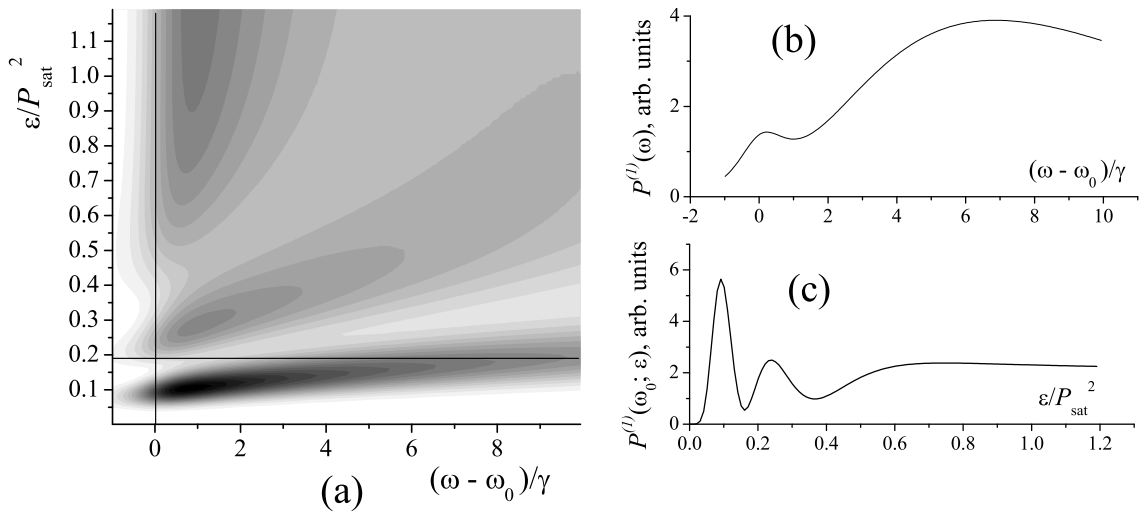}
  \caption{
  (a) Four-wave mixing spectrum corresponding to $2\mathbf{k}_1 - \mathbf{k}_2$
  as a function of the normalized pulse area. The parameters of the
  system are chosen to be $\beta'/\gamma = 5\cdot 10^2$, $\beta'/2\beta'' =
  5$. The horizontal and the vertical lines show the positions of the sections of the
  surface $P^{(1)}(\omega; \epsilon)$ presented in (b) and (c).
  (b) The FWM spectrum at the fixed value $\epsilon/P_{sat}^2 = 0.19$.
  (c) The dependence of the FWM response at the exciton frequency
  $P^{(1)}(\omega_0; \epsilon)$ as a function of the pulse area.}\label{fig:typical_spectrum}
\end{figure*}

Using the solutions for the free and driven polarization in
Eqs.~(\ref{eq:effective_linear}) and (\ref{eq:immediate response}), we
solve the time evolution of the polarization in all the four regions $t_-'
< t < t_+'$, $t_+' < t < t_-''$, $t_-'' < t < t_+''$, and $t > t_+''$.
Assuming that the system initially is in the ground state $P(t_-')\equiv0$
we find the exciton polarization created by the first pulse
\begin{equation}\label{eq:first_pulse_response}
  P_1(t_+';\xx) = -i e^{i \mathbf{k}_1\cdot \xx}P_1(t_+'),
\end{equation}
where
\begin{equation}\label{eq:amplitude_first_response}
 P_1(t_+') = P_{sat} \tanh\left(\frac{{\epsilon}}{P_{sat}}\right).
\end{equation}
It follows from this equation that the saturation effect for the first pulse
reduces to a simple (although nonlinear) renormalization of the signal area.
Using Eq.~(\ref{eq:first_pulse_response}) in Eq.~(\ref{eq:p_amp_solution})
we can see that the effective frequency $\Omega(t; \xx)$ determining the
dynamics of the polarization is constant
across the sample. As a result there is no coupling between the 
exciton modes characterized by different $\mathbf{k}$.

Denoting the delay time, the time separation between the pulses, by
$\tau=t_-''-t_+'$ we obtain the polarization right before the arrival of the
second pulse
\begin{equation}\label{eq:second_initial_condition}
  P_{10}(t_-''; \xx) = P_1(t_+';\xx) \theta_{10}(\tau),
\end{equation}
where $\theta_{10}(\tau) = \theta(\tau; |P_1(t_+')|)$. The spatial
distribution of the polarization $P_{10}(t_-''; \xx)$ plays the role of the
initial condition for the immediate response with respect to the second
pulse according to Eq.~(\ref{eq:immediate response}). For the analysis of
the time evolution of the polarization during the second pulse it is
convenient to factor $\exp(i\mathbf{k}_2\cdot \xx)$ out of $P_2(t;\xx)$
introducing
\begin{equation}\label{eq:factorized_P}
  \widetilde{P}_2(t; \Delta \mathbf{k}\cdot \xx) = e^{-i\mathbf{k}_2\cdot
  \xx} P_2(t; \xx)
\end{equation}
with $\Delta\mathbf{k} = \mathbf{k}_1 - \mathbf{k}_2$. The reduced
distribution $\widetilde{P}_2(t; \Delta \mathbf{k}\cdot \xx)$ satisfies
Eq.~(\ref{eq:eq_of_motion}) with the modified spatial profile of the
external field $\widetilde{E}(t) = e^{-i\mathbf{k}_2\cdot \xx}E(t;\xx)$. The
form of the initial conditions in Eq.~(\ref{eq:immediate response}) changes
according to
\begin{equation}\label{eq:initial_conditions_new}
\widetilde{P}_{10}(t_-''; \Delta \mathbf{k}\cdot \xx) = -i\theta_{10}(\tau)P_1(t_+')
e^{i\Delta\mathbf{k}\cdot \xx}.
\end{equation}

It follows from Eqs.~(\ref{eq:immediate response}) and
(\ref{eq:initial_conditions_new}) that for the second pulse the role of
the saturation effect is two-fold. It modifies the pulse area and excites
all modes $\mathbf{k}^{(n)}$ rather than just a single mode as we had for
the first pulse. We present the polarization as a superposition of the
multi-wave mixing modes
\begin{equation}\label{eq:superposition_modes}
  \widetilde{P}_2(t; \kappa) = \sum_n {P}_2^{(n)}(t)
  e^{in \kappa},
\end{equation}
where $\kappa = \Delta\mathbf{k}\cdot \xx$ and $P_2^{(n)}(t)$ are the
amplitudes of the multi-wave mixing polarizations. Substituting this
representation into Eq.~(\ref{eq:effective_linear}) we can see that in terms
of the multi-wave mixing modes the free dynamics of the polarization can be
presented as the dynamics of a system with an infinite number of degrees of
freedom coupled to each other due to the nonlinearity
\begin{equation}\label{eq:binding_model}
  \frac{d}{dt}{P_2}^{(n)}(t) = -(i\omega_0 + \gamma)P_2^{(n)}(t) -
  i\sum_{m}\Omega_{n-m}(t) P_2^{(m)}(t),
\end{equation}
where $ \Omega_{n}(t) = \beta(2\pi)^{-1}\int_{-\pi}^\pi
|\widetilde{P}_2(t; \kappa)|^2 e^{-i\kappa n}d\kappa$.
The initial conditions for Eqs.~(\ref{eq:binding_model}) are constituted
by the immediate response to the second pulse thanks to the saturation
effect. This picture clearly illustrates the difference between the effect
of the Pauli blocking and the Coulomb interaction on formation of the
multi-wave mixing response.

Using the solution of the equation of motion for free polarization
dynamics we find
\begin{equation}\label{eq:final_solution}
 P_{2}^{(n)}(t) = \frac{1}{2\pi} \int_{-\pi}^\pi \widetilde{P}_2(t_+''; \kappa)
 \theta\left(t; |\widetilde{P}_2(t_+''; \kappa)|\right)e^{-i\kappa n} d\kappa.
\end{equation}
This equation and Eqs.~(\ref{eq:immediate response}) and
(\ref{eq:initial_conditions_new}) give the exact evolution of the exciton
polarization in the limit of short excitation pulses in the two-pulses
scheme. Formally, one can obtain from Eq.~(\ref{eq:final_solution}) the
spectrum $P_{2}^{(n)}(\omega)$ using the Fourier transform of
$\theta(\tau; |\widetilde{P}_2(t_+''; \kappa)|)$ with respect to time
\begin{equation}\label{eq:pi_transform}
  \theta(\omega; |\widetilde{P}_2(t_+''; \kappa)|) = \frac{i}{\gamma(w + i/2)}\,
  {}_2 F_1\left(1; \frac{1}{2}+iX; \frac{3}{2} - iw;
  -\frac{\beta''|\widetilde{P}_2(t_+''; \kappa)|^2}{\gamma}\right),
\end{equation}
where $w=(\omega - \omega_0)/2\gamma$, $X=\beta'/2\beta''$ and ${}_2 F_1$
is the hypergeometric function. Technically, however, because the second
argument of this function is a complex number it may be more efficient to
calculate the spectrum using the time series $P_{2}^{(n)}(t)$.

The spectrum corresponding to the four-wave mixing directions
$2\mathbf{k}_2 - \mathbf{k}_1$ and $2\mathbf{k}_1 - \mathbf{k}_2$ is
obtained from Eq.~(\ref{eq:final_solution}) taking $n=-1$ and $n=2$,
respectively, and for $n=2$ is shown in Fig.~\ref{fig:typical_spectrum} as
a function of the pulse area. In order to estimate the relation between
the decay rate and the non-linear parameter we have used the expressions
following from the microscopic consideration\cite{TAKAYAMA:2002} $\beta'
\approx 1.52 a_B^2 E_b $ and $P_{sat}^2 = 7/4\pi a_B^2$, where $E_b$ is
the exciton Rydberg and $a_B$ is the exciton Bohr radius.

The spectrum has two typical features. The first one consists of branches
detached from the exciton frequency with increasing pulse area. These
branches for sufficiently high amplitudes of the excitation field may
manifest themselves on the spectrum in the form of multiple resonances.
The second interesting feature is the oscillatory character of the field
dependence of $|P_{2}^{(2)}(\omega_0;\epsilon)|^2$, that is the FWM
response at the exciton frequency.

\subsection{The FWM spectrum in the limit of negligible EID}

We start the discussion of these features from the simplest case
$\beta''=0$ (vanishing EID) and high $P_{sat}$ (low saturation regime).
The first assumption simplifies the effect of the initial conditions on
the polarization dynamics while the second simplifies the relation between
the excitation field and the polarization of the immediate response, so
that
\begin{equation}\label{eq:immediate_response_low_saturation}
  \widetilde{P}_2(t_+''; \kappa) = -i\left(|{\epsilon}_1\theta_{10}(\tau)|e^{-i\kappa+i\phi_1}
  + |{\epsilon}_2|e^{i\phi_2}\right),
\end{equation}
where $\phi_1 = \arg({\epsilon}_1\theta_{10}(\tau))$ and $\phi_2 =
\arg({\epsilon}_2)$.

It is interesting to note that according to Eqs.~(\ref{eq:binding_model})
and (\ref{eq:p_amp_solution}) in this approximation since only
$\Omega_{\pm1}(t)$ differ from zero the dynamics of the polarization is
described by the equation of motion for a 1D tight-binding model with the
time-dependent coupling between neighboring sites $\propto
\beta|\epsilon_2\epsilon_1 \theta_{10}(\tau)| e^{-2\gamma t}$. The
relation between the amplitudes of the excitation at different sites gives
the relation between the amplitudes of the signals corresponding to
multi-wave mixing. Initially the excitation is localized on sites $n=0,1$
and with time it propagates along the chain giving
\begin{widetext}
\begin{equation}\label{eq:solution_simplest}
\begin{split}
 P_{2}^{(n)}(t) = -e^{in(\phi_2 - \phi_1 - \pi/2)+i\phi_2}
 \exp\left\{-i\omega_0 t - \gamma t - \frac{i\beta I}{2\gamma}\left(1 - e^{-2\gamma t}\right)\right\}
 \times
 \\
 \left\{|\epsilon_1 \theta_{10}(\tau)| J_{n-1}\left[\eta(1 - e^{-2\gamma t})\right]
 + i|\epsilon_2| J_{n}\left[\eta(1 - e^{-2\gamma t})\right]
 \right\},
\end{split}
\end{equation}
where $J_n$ are the Bessel functions of the first kind, $I = |\epsilon_1
\theta_{10}(\tau)|^2 + |\epsilon_2|^2$, and $\eta = \beta
|\epsilon_2\epsilon_1 \theta_{10}(\tau)|/\gamma$. Deriving
Eq.~(\ref{eq:solution_simplest}) we have used the Jacobi-Anger
expansion\cite{Arfken_Weber} $\exp(i z \cos \kappa) =
\sum_{n=-\infty}^\infty i^n J_n(z) e^{in\kappa}$.
\end{widetext}

\begin{figure}
  \includegraphics[width=3in]{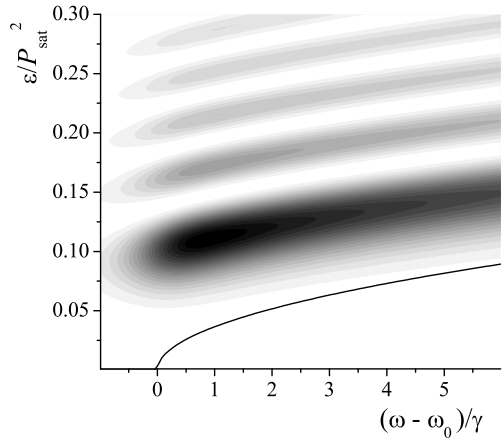}
  \caption{The form of the four-wave mixing spectrum in low saturation regime
  in the case of negligible EID ($\beta''=0$) and $\beta'/\gamma = 2.5\cdot 10^2$.
  The solid line shows the position of the resonant frequency
  determined by $\omega = \omega_0 + \beta I + \eta\gamma$.
  }\label{fig:spectrum_simple}
\end{figure}

The spectrum of the four-wave mixing signal corresponding to $n=2$ is
shown in Fig.~\ref{fig:spectrum_simple}. Shortly after the excitation, for
$t \ll 1/2\gamma$, the amplitude of the multi-wave mixing signal drops
exponentially with the order of mixing, $\sim (\eta \gamma t/2)^n$. The
exponential drop holds asymptotically in time if $\eta < 1$. This result
agrees with the perturbational approach. The situation, however,
drastically changes if the parameters of the system are such that $\eta
\gg1$. In this case, the intensity of the multi-wave mixing signal becomes
independent of its order starting time $t\gtrsim-\ln(1-\eta^{-1})/2\gamma
\approx 1/2\beta |\epsilon_2\epsilon_1 \theta_{10}(\tau)|$. This
consideration suggests naturally to identify $\eta = 1$ as a critical
value that separates perturbative and non-perturbative regimes.

The detailed form of the spectrum is essentially determined by the fact
that the effective coupling between the modes vanishes with time. Right
after arrival of the second pulse the polarization oscillates with the
frequency detuned from the exciton frequency by the value $\sim \beta I +
\eta\gamma$, where the second term estimates the contribution of the
Bessel functions. This detuned frequency qualitatively describes the
dependence of the frequencies of the detached resonances on the signal
area (see the bold solid line in Fig.~\ref{fig:spectrum_simple}). In
particular it shows that the resonances in the spectrum widen with
increasing nonlinear parameter $\beta$. In the opposite limit, $t \gg
1/\gamma$, the coupling between the modes vanishes and, as a result, the
exciton polarization oscillates with the non-modified exciton frequency
$\omega_0$. These oscillations give rise to the resonant behavior at the
exciton frequency, for which dependence on the pulse area is determined by
the asymptotic value of the Bessel functions $J(\eta)$. Therefore, the
response at the exciton frequency essentially depends on whether the
system is in perturbative or non-perturbative regime. We illustrate the
difference between these regimes considering the effect of the positive
and negative delay time. The expression for the negative delay time can be
obtained from Eq.~(\ref{eq:solution_simplest}) by considering the signal
in the ``conjugate" direction $\bar{n} = 1-n$ and exchanging $\epsilon_1
\leftrightarrow \epsilon_2$.

In the perturbative regime, $\eta \ll1$, we obtain
\begin{equation}\label{eq:perturb_positive_negative}
\begin{split}
 \left|P_{2}^{(n)}(\omega_0; \tau>0)\right|^2 \sim 4\eta_0^{2n-2}
 \left|\epsilon_1 n \theta_{10}^n(\tau)\right|^2, \\
 \left|P_{2}^{(n)}(\omega_0; \tau<0)\right|^2 \sim 4\eta_0^{2n-2}
 \left|\epsilon_1 n \theta_{10}^{n-1}(-\tau)\right|^2,
\end{split}
\end{equation}
where $\eta_0 = \beta |\epsilon_1\epsilon_2|/\gamma$. In this regime
$|\theta_{10}(\tau)| \approx \exp(-\gamma \tau)$, thus, the positive delay
signal decays with time constant $2n\gamma$ while for the negative
delay it decays more slowly with the constant $2(n-1)\gamma$. For the case
of FWM signal this result
agrees with the perturbational calculations.\cite{WEGENER:1990}

In the non-perturbative regime the positive and negative delay signals are
determined by the oscillating asymptotics of the Bessel
functions\cite{Gradshteyn}
\begin{equation}\label{eq:nonpert_positive_negative}
\begin{split}
 \left|P_{2}^{(n)}(\omega_0)\right|^2 \sim & \frac{1}{\pi \eta}
 \left[\left|\epsilon_1 \theta_{10}(\tau)\right|^2 + \left|\epsilon_2\right|^2 \right. \\
 &\left.\mp (-1)^n
 \left(\left|\epsilon_1 \theta_{10}(\tau)\right|^2 - \left|\epsilon_2\right|^2\right)
 \sin(2\eta)
 \right],
\end{split}
\end{equation}
where ``$-$" (``$+$") sign corresponds to the positive (negative) delay.
Writing Eq.~(\ref{eq:nonpert_positive_negative}) we have neglected the
oscillating term $\propto \cos(2\eta)$ vanishing as $(\left|\epsilon_1
\theta_{10}(\tau)\right|^2 + \left|\epsilon_2\right|^2)/\eta^2$ with
increasing signal area. Thus, for both positive and negative delays the
response at the exciton frequency saturates at the oscillations with the
period $\eta_T = 2\pi$. The strong asymmetry between these cases specific
for the perturbative regime does not hold any longer and the only
difference is the phase of the oscillations. It should be noted that in
Ref.~\onlinecite{SHACKLETTE:2003} the saturation of the FWM response was
attributed to the renormalization of the pulse area by the EID and EIS.
The present consideration, however, suggests that the origin of the
non-trivial dependence of the response on the pulse area is the
redistribution of the excitation over multi-wave mixing modes. Considering
the identity $\sum_n J_n^2(z) = 1$ for the limits $z\ll1$ and $z\gg1$ it
can be seen that such redistribution is especially effective in the
nonperturbative regime resulting in essential suppression of the FWM
response.

It should be emphasized that the mechanism of the oscillatory dependence
of the response at the exciton frequency is different from the Rabi
oscillations,\cite{BINDER:1990,OSTREICH:1993}
which would correspond to the non-monotonous dependence of the immediate
response on the excitation field. In the case under consideration the
oscillations are the result of the free dynamics of the exciton
polarization when the external field is turned off. The physics of the
Rabi oscillations and of the oscillations of $|P^{(n)}_{2}(\omega;
\eta)|^2$, of course, are essentially the same. As noted above the
dynamics of the polarization in the case under consideration appears
analogous to a 1D tight-binding model with vanishing coupling between the
neighboring sites. In quantum mechanical terms it can be described as a
multiple level system, where the levels correspond to
the exciton 
modes, with time dependent field $V_{ij}(t)$, which couples different
levels. Depending on the ``area" of the off-diagonal elements, $\int dt
V_{ij}(t)$, one has the oscillations of the final populations of the
different levels. Translated to the language of the multi-mixing signals
$P_{2}^{(n)}$ this result implies the oscillations of
$|P_{2}^{(n)}(\omega_0)|^2$ since asymptotically, as has been noted, one
has the dynamics determined by the non-perturbed exciton frequency.

\subsection{The effect of the excitation induced decay}

In order to study the effect of the EID on the spectrum (compare
Figs.~\ref{fig:typical_spectrum} and \ref{fig:spectrum_simple}) we use
Eq.~(\ref{eq:pi_transform}) assuming that $|{\epsilon}_1
\theta_{10}(\tau)| = |{\epsilon}_2| = {\epsilon}$. Considering the
asymptotic values of the hypergeometric function\cite{Abramowitz} in the
limit $\eta''=2\beta''{\epsilon}^2/\gamma \gg 1$ we can approximately
present the spectrum in the form
\begin{equation}\label{eq:factorized_spectrum}
  P_{2}^{(n)}(\omega) \approx \frac{\epsilon}{\sqrt{\eta''}}\left[
  C(\omega; X)A_2 e^{-iX \ln \eta''}+\frac{1}{\eta''}\,
  \frac{A_1}{iX-1/2}\right],
\end{equation}
where
\begin{equation}\label{eq:C_def}
  C(\omega; X) = \frac{\Gamma(1/2-iw)\Gamma(1/2-iX)}{\Gamma(1-iw-iX)}
\end{equation}
and $A_1=A_n(1)$, $A_2=A_n(1/2)$ with $A_n(p)$ depending on $X$ only
\begin{equation}\label{eq:def_Anp}
\begin{split}
  A_n(p) = \frac 1{2\pi}\int_{-\pi}^\pi & d\kappa (1+e^{-i\kappa})e^{-in\kappa} \\
  & \times \exp\left\lbrace -(iX+p)\ln\left[2\cos^2\left(\frac{\kappa}{2}\right)\right]\right\rbrace.
\end{split}
\end{equation}

Similarly to the case of negligible EID, the response oscillates and
reaches the saturation in the high excitation limit $\eta''\gg1$. At the
exciton frequency the magnitude of the signal is
\begin{equation}\label{eq:EID_limit}
  \left|P_{2}^{(n)}(\omega_0)\right|^2 \sim \frac{\pi\epsilon^2}{X\eta''}  \left|A_2 \right|^2
 \tanh(\pi X).
\end{equation}
This saturation also is of the dynamical origin since we work in the
regime of weak saturation due to the Pauli blocking ($\epsilon/P_{sat}^2
\ll 1$). The significant difference with the previously considered
situation is that now the minima of the spectrum as a function of the
pulse area are not equally spaced as the oscillating part has the form
$\propto \cos(X \ln \eta'')$. Thus, with the decreasing ratio between the
real and imaginary parts of the non-linear parameter the crossover from
linear to logarithmic scale occurs.

Finally, we discuss the effect of the saturation parameter. Qualitatively,
this effect can be understood as follows. The non-monotonous behavior of
the semiconductor response on the pulse area studied above is supported by
unrestricted increase of the polarization of the immediate response [see
Eq.~(\ref{eq:immediate_response_low_saturation})]. However, the saturation
effect renormalizes the pulse area so that the magnitude of the
polarization can not exceed $P_{sat}$. For example, from the perspective
of the discussion of the effect of EID this means that
Eq.~(\ref{eq:factorized_spectrum}) remains valid only if two restrictions
are met $2\beta''{\epsilon}^2/\gamma \gg 1$ and ${\epsilon} < P_{sat}$.
This imposes the restriction of the decay rate to be sufficiently small
$\gamma \ll \beta'' P_{sat}^2$. 


\section{Conclusion}

We have studied non-perturbative effects in four-wave mixing spectra of
semiconductors. These effects are analyzed using the exact solution of the
non-linear equation of motion of the exciton polarization taking into
account excitation induced shift (EIS), excitation induced decay (EID) and
the saturation effect phenomenologically. We found that the interaction
between the excitons accounted by EIS  leads to two specific spectral
features --- a split of the exciton peak and a non-monotonous dependence
of the response at the exciton frequency $\omega_0$  on the magnitude of
the external field. The important characteristic of the splitting is that
new spectral features should appear at frequencies higher than $\omega_0$.
This allows one to make a distinction between the effect of interaction of
the 
exciton modes and the manifestation of bound biexciton
states, which should modify the spectrum at frequencies lower than $\omega_0$.

We would like to emphasize that these effects do not appear in any order of
the perturbational ($\chi^{(n)}$) approach. It can be shown that the
appearance of additional spectral features can be traced as a divergence of
the perturbational series. It should be stressed out that the crossover from
the perturbative to non-perturbative regimes is governed by parameters that
essentially depend on the decay rate (e.g. $\beta
|{\epsilon}_1\theta_{10}(\tau){\epsilon}_2|/\gamma \sim 1$ in the case
$\beta''=0$). This means that the condition of low excitation itself does
not necessarily warrant the validity of the perturbation theory. As an
ultimate example one can consider the model with $\gamma = 0$ when the
spectrum (for $\tau =0$) has the form $P(\omega)\propto 1/\sqrt{(\omega -
\omega_0 - \beta I)^2 - \beta^2 I^2}$ with the exciton peak being splitted
for arbitrary low excitations.

\acknowledgments

We would like to acknowledge the support for this work from the National
Science Foundation under grant number ECCS-0725514 and through the DARPA/MTO
Young Faculty Award under Grant No. HR0011-08-1-0059.


\end{document}